\documentclass[a4paper,12pt]{article}
\pdfoutput=1
\usepackage{cite}
\usepackage{jheppub}
\usepackage{hyperref}
\hypersetup{colorlinks=true}
\usepackage[T1]{fontenc} 
\title{Neumann-Rosochatius system for rotating strings in $AdS_3 \times S^3\times S^3\times S^1$ with flux}
\author{Adrita Chakraborty$^1$, Rashmi R. Nayak$^2$, Priyadarshini Pandit$^1$, Kamal L. Panigrahi$^1$}
 \affiliation{$^1$ Department of Physics, Indian Institute of Technology Kharagpur-721302, India}
 \affiliation{$^2$Centre for Ocean, River, Atmosphere and Land Sciences (CORAL), Indian Institute of Technology Kharagpur- 721302, India}
\emailAdd{adimanta09@iitkgp.ac.in}
\emailAdd{rashmi@coral.iitkgp.ac.in}
\emailAdd{pandit006@iitkgp.ac.in}
\emailAdd{panigrahi@phy.iitkgp.ac.in}
\abstract{Strings on $AdS_3 \times S^3\times S^3\times S^1$ with mixed flux exhibit exact integrability. We wish to construct an integrable Neumann-Rosochatius (NR) model of strings starting with the type IIB supergravity action in $AdS_3 \times S^3\times S^3\times S^1$ with pure NSNS flux. We observe that the forms of the Lagrangian and the Uhlenbeck integrals of motion of the considered system are NR-like with some suitable deformations which eventually appear due to the presence of flux. We utilize the integrable framework of the deformed NR model to analyze rigidly rotating spiky strings moving only in $S^3\times S^1$. We further present some mathematical speculations on the rounding-off nature of the spike in the presence of non-zero angular momentum $J$ in $S^1$.}

\begin{document}
\maketitle
	\flushbottom
\section{Introduction}
 The strongly coupled non-perturbative behaviour of gauge theories is considered to be one of the most tangled long-standing problems in the field of fundamental research. Since the last two decades, the duality  \cite{Maldacena:1997re,Witten:1998qj,Gubser:1998bc} between lower dimensional strongly coupled conformal field theory and higher dimensional weakly coupled gravity theory in various Anti-de Sitter (AdS) spacetimes has been greatly felicitated in removing the problem of non-perturbative behaviour of gauge theories, in the planar limit. The most studied example of the AdS/CFT correspondence is the duality between $\mathcal{N}=4 $ Super Yang-Mills (SYM) theory and type IIB superstrings in $AdS_5\times S^5 $ background. However, in the past few years, numerous attempts has been taken in order to establish the duality in different sub-sectors of the theories on both sides of the correspondence. Semi-classical approximation being one of the most appropriate method used extensively in order to find the spectrum in different target space geometries. In this method we find the dispersion relation between various conserved charges of semi-classical string dynamics in $AdS$ space which in turn gives the anomalous dimensions of the dual boundary field theory operators \cite{GKP,Berenstein:2002jq,Frolov:2003qc,Tseytlin:2003ii,Engquist:2003rn,spiky1}. In particular, the study of rigidly rotating string in the large charge limit moving in some curved background in order to establish the AdS/CFT holographic realization has gained much attention \cite{Kruczenski:2008bs,Ishizeki:2008tx,Jevicki:2008mm,Freyhult:2009bx,Kruczenski:2010xs,Banerjee:2015nha,Losi:2010hr,Banerjee:2019puc,Biswas:2020wfn}. In order to have a better grasp on the spectrum of $AdS_5\times S^5$ superstring theories, various other class of classical string solutions has also been studied to a great extent.\\ 
 In establishing the AdS/CFT correspondence, planar integrability has also played one of the most significant role on both side of the correspondence. The first evidence in connection with this was found in \cite{Minahan:2002ve} where the anomalous dimensions of one loop dilatation operators in the SU(2) sector of planar $\mathcal{N}=4$ SYM were matched with the eigenvalues of diagonalised Hamiltonian acting upon the integrable Heisenberg XXX spin chain \cite{Beisert:2003jj,Beisert:2003yb}. Pioneered by this observation, integrability helps in learning more powerful spectrum of the local operators in a vastly simplified manner. The notion of integrability for the holographically celebrated $AdS_3/CFT_2$ was initiated from \cite{Babichenko:2009dk}. Conventionally, there occur two plausible integrable set-ups in the case of $AdS_3/CFT_2$ \cite{Sfondrini,SST} namely the type IIB GS string construction in $AdS_3\times S^3\times \mathcal{M}^4$, $\mathcal{M}^4\equiv T^4$ \cite{Ads3S3T4A,BianchiHoare,BothM4,completeworldsheet,BMN mismatch,BPS} or $S^3\times S^1$ \cite{Rughoonauth,Borsato:2015mma,Abbott,Abbott:2013mpa,EberhardtGopakumar,Tong,Pittelli:2017spf,wulff,review}. The underlying 
$\mathbb{Z}_4$ automorphism in these background $AdS_3\times S^3_1\times S^3_2\times S^1$ represents a one parameter family of inequivalent geometries that are labelled by the radii of  the two sphere. Here these radii of curvature of the two spheres are conveniently expressed in terms of a family of one parameter $\alpha$ as\cite{Gauntlett:1998fz},
\begin{equation}
    \alpha\equiv \cos^2{\varphi}= \frac{R_3^2}{R_1^2}=1-\frac{R_3^2}{R_2^2},~~\frac{R_3^2}{R_2^2}\equiv \sin^{2}\varphi
    \label{triangle}
\end{equation} where $R_1,R_2$ and $R_3$ are the radii of curvature for the geometries $S^3_1$, $S^3_2$ and $AdS_3$ respectively. The parameter $\alpha$ assumes values within $0<\alpha<1$ when a suitable reality condition is imposed on the exceptional Lie super-algebra $\mathfrak{d}\left(2,1,;\alpha\right)$ which satisfies exactly the same triangle equality (\ref{triangle}) for its invariant bilinear forms. For this range of $\alpha$, the bosonic sector requires the action to be that of a sigma model in a target-space with group manifold $SU(1, 1)\times SU(2)\times SU(2)$ which is nothing but $AdS_3\times S^3\times S^3$. However, there are two special limits for $\alpha$. One is $\alpha\rightarrow 1$ implying that radius of one of the spheres blows up and that sphere being a flat space gets compactified on a $T^3$, eventually representing the usual $AdS_3\times S^3\times T^4$ background. In this case, the symmetry algebra shrinks up to some abelian factors to $PSU(1, 1|2)\times PSU(1, 1|2)$. The other limit demands $\alpha=\frac{1}{2}$, i.e., $\phi=\frac{\pi}{4}$ with which the spheres $S^3_1$ and $S^3_2$ become identical and the algebra merges into the classical $\mathfrak{osp}(4|2)$ superalgebra.

The existence of classical integrability of the $AdS_3\times S^3_1\times S^3_2\times S^1$ was observed in \cite{Sundin:2012gc} in the presence of mixed flux. Also, there are some recent findings on the integrability of the same by using the integrable worldsheet S-matrix realizations, both for pure RR and generalised mixed flux \cite{Borsato:2012ud}. However, it is crucial to note that unlike the $AdS_3\times S^3\times T^4$ background, we still do not have a clear duality in $AdS_3\times S^3_1\times S^3_2\times S^1$ background. Reason behind such discrepancy is the lack of sufficient consistency of supergravity approximation. For pure NSNS case, various backgrounds having an $AdS_3$ factor consist of WZW prescription with explicitly solvable $\mathfrak{sl}(2,\mathbb{R})$ symmetry algebra has been studied. String theory on $AdS_3\times S^3_1\times S^3_2\times S^1$ with pure NSNS flux can possess such WZW model description which involves $\mathcal{N}=1$ affine algebras $\mathfrak{sl}(2)_k\oplus\mathfrak{sl}(2)_{k_1}\oplus\mathfrak{su}(2)_{k_2}\oplus\mathfrak{u}(1)$\cite{deBoer:1999gea}. Here the relevant levels of the algebras are also related to the parameter $\alpha$ of the relative spherical geometries and the relation goes as
\begin{equation*}
    \alpha=\frac{k}{k_1}=\cos^2{\varphi},~~ 1-\alpha=\frac{k}{k_2}=\sin^2{\varphi}
\end{equation*}The corresponding dual gauge theory can thus be explored with the help of the underlying chiral representation theory. In \cite{Gukov:2004ym}, the protected BPS spectrum achieved with pure NSNS flux has an indication toward the large $\mathcal{N}=(4,4)$ superconformal symmetry only when the charge of the fivebrane wrapped around one of the 3-spheres becomes unity. This is retrieved by using the relation (\ref{triangle}).

Our present work is devoted to the formulation of one-dimensional integrable NR Model in the geometry of $AdS_3\times S^3\times S^3\times S^1$ supported by pure NSNS flux and having the fundamental string as the probe. The NR model, as an integrable extension of earlier Neumann model, elucidates the constrained motion of a harmonic oscillator on a unit sphere under the influence of an extra centrifugal potential of the form $\frac{1}{x_i^2}$. Being one of the earliest classical integrable model, NR model has been studied as an exemplary ground to compute for large class of  string solutions in various integrable 10d $AdS_d\times X^{10-d}$ backgrounds with or without flux by means of some generic NR ansatz \cite{Chakraborty:2020las,Chakraborty:2019gmt,Hernandez:2017raj,Arutyunov:2016ysi,Hernandez:2015nba,Ahn:2008hj,Hernandez:2014eta}.
The appearance of NR model is approved to be quite robust for many deformed backgrounds as well as for some probe strings other than fundamental ones.
We explored here, the NR construction by assuming the fundamental string as a probe and explicitly deriving the Lagrangian, Hamiltonian and Uhlenbeck constants. The suitable values of flux parameters are calculated accordingly from the equations of motion of the type IIB supergravity and the equation (\ref{triangle}). 

The organisation of the article is as follows. In section 2 we revisit the $AdS_3\times S^3\times S^3\times S^1$ supergravity solution with pure NSNS flux and use various parametric relation along with the $\frac{1}{2}$-BPS supersymmetry preserving relation between the radii of curvature, to verify whether the flux parameters assume values consistent with those for pure NSNS case. The geometrical constraints as well as the rotating string ansatz are also presented in this section. Section 3 includes the Lagrangian and Hamiltonian formulation for the NR Model along with the construction of Uhlenbeck integrals of motion. We devote section 4 to the study of a class of rotating string with explicit derivation of rigidly rotating $N$-spike string profile moving in $S^3\times S^3$ \cite{Kruczenski:2006pk} in the presence of non-zero angular momentum $J$ and winding number $m$ in $S^3$ \cite{Nayak:2021hrd} by solving the integrable NR model Lagrangian. Here we mathematically speculate the rounding off nature of the spike appearing explicitly due to the additional angular momentum $J$ in $S^1$, unlike the general spiky string profile where the spikes end in cusps. We conclude in the last section with a brief future outlook.\\

\section{Gravity dual Background for $AdS_3 \times S^3 \times S^3$ with mixed flux}
The most general form of the bosonic part of  type IIB Supergravity action in the string frame with the fermionic terms dropped consistently, is given by ,
\begin{equation}
   \begin{split}
        S_{IIB}=\frac{1}{2\kappa^2_{10}}\Big[ \int d^{10}x \sqrt{-g}&\Big(e^{-2\phi}\left(R+4\partial_M\phi\partial^M\phi-\frac{1}{2}|H_3|^2\right)-\frac{1}{2}|F_{(1)}|^2-\frac{1}{2}|F_{(3)}|^2\\&-\frac{1}{4}|F_{(5)}|^2\Big) -\frac{1}{2}\int C_{(4)}\wedge H_{(3)}\wedge F_{(3)}\Big]
        \end{split}
\end{equation}
    where $\kappa_{10}^2=8\pi G_{10}$ with $G_{10}$ being 10D Newton's gravitational constant, $H_3$ and $F_{p+2}$ are respectively the 3-form NSNS and $(p+2)$-form RR fluxes present in the background. These fluxes may be expressed as,
\begin{equation}
 H_3=dB_2,~~~F_{p+2}=dC_{p+1},~~~ p=-1,1,3,
 \label{fluxstrength}
\end{equation}with $B_2$ and $C_{p+1}$ are 2-form NSNS and $(p+1)$-form RR fluxes respectively. In the above equations, the fluxes follow the general expression
\begin{equation}
    |K_p|^2=\frac{1}{p!}G^{M_1N_1}G^{M_2N_2}.....G^{M_pN_p}K_{M_1...M_P}K_{N_1...N_p}=\frac{1}{p!}K_p^2,
    \label{field}
\end{equation}
where, $K_p$ is a $p$-form field. Here we solve type IIB equations of motion for $AdS_3\times S^3\times S^3\times S^1$ background with mixed RR and NS-NS  fluxes, where the subscript 1 and 2 on $S^3$ represents two 3 spheres with different radii, $R_1$ and $R_2$ respectively.
 
   
The metric for $AdS_3\times S^3_1\times S^3_2\times S^1$ in terms of global coordinates can be written as
\begin{equation}
    ds^2=ds^2_{AdS_3}+ds^2_1+ds^2_2+dw^2,
\end{equation}where $w$ is the coordinate along $S^1$ and 
\begin{subequations}
\begin{align}
   &ds_{AdS_3}^2=\frac{R_3^2}{x_2^2}\left(-d\tau_p^2+dx_1^2+dx_2^2\right)\label{Ads3},\\&
   ds_1^2=R_1^2\left[d\beta_{1}^2+\sin^2{\beta_1}d\beta_2^2+\cos^{2}{\beta_1}d\beta_3^2\right],\\&
   ds_2^2=R_2^2\left[d\beta_{4}^2+\sin^2{\beta_4}d\beta_5^2+\cos^{2}{\beta_4}d\beta_6^2\right],
   \end{align}
\end{subequations}respectively represent the metric for a coordinate patch that covers a part of the $AdS_3$ spacetime and the round metrics of the two 3-spheres present in the background. By using the coordinate transformation as
\begin{subequations}
\begin{align*}
    &\tau_p=\frac{R_3\cosh{\rho}\sin{t}}{\sinh{\rho}\sin{\phi}-\cosh{\rho}\cos{t}},\\& x_1=\frac{R_3\sinh{\rho}\cos{\phi}}{\sinh{\rho}\sin{\phi}-\cosh{\rho}\cos{t}},\\&x_2=\frac{R_3}{\sinh{\rho}\sin{\phi}-\cosh{\rho}\cos{t}}
    \end{align*}
\end{subequations}in (\ref{Ads3}), a convenient form of the  metric of full $AdS_3$ spacetime may be obtained in terms of global cylindrical coordinates $(t,\rho,\phi)$ as \begin{equation}
    ds^2_{AdS_3}=R_3^2\left[-\cosh^2{\rho}dt^2+d\rho^2+\sinh^2{\rho}d\phi^2\right].
\end{equation}We will continue our work with the global cylindrical $AdS_3$ metric. The associated pure NSNS 3-form background flux is considered as
\begin{equation}
H=b_0\omega_0+b_1\omega_1+b_2\omega_2,
    \label{NSNS}
\end{equation}with volume fluxes
\begin{subequations}
\begin{align}
    &\omega_0=vol(AdS_3)=R_3^3\cosh{\rho}\sinh{\rho}dt\wedge d\rho\wedge d\phi,\\&
    \omega_1=vol(S^3_1)=R_1^3\sin{\beta_1}\cos{\beta_1}d\beta_1 \wedge d\beta_2 \wedge d\beta_3,\\&
    \omega_2=vol(S^3_2)=R_2^3\sin{\beta_4}\cos{\beta_4}d\beta_4 \wedge d\beta_5 \wedge d\beta_6,
    \end{align}
\end{subequations} where $b_i$'s denote the flux coefficients satisfying the constraint
\begin{equation}
    \frac{1}{6}H_{MNP}H^{MNP}=-{b}_0^2+b_1^2+b_2^2,
    \label{parameter}
\end{equation}
followed from the equation of motion of $\phi$, where $\phi$ is taken to be zero.\\
The Ricci scalar is\\
\begin{equation*}
  \mathcal{R}=6\left( \frac{1}{R_1^2}+\frac{1}{R_2^2}-\frac{1}{R_3^2}\right) .
\end{equation*}
    Using equation (\ref{fluxstrength}) the nonzero pure NSNS 2-form flux components associated to the desired background are calculated from (\ref{NSNS}) as
    \begin{subequations}
\begin{align}
    &B_{t\phi}=b_0 R_3^2\sinh^2{\rho},\\&
    B_{\beta_2\beta_3}=-b_1R_1^2\cos^2{\beta_1},\\&
    B_{\beta_5\beta_6}=-b_2R_2^2\cos^2{\beta_4}.
   \end{align}
    \end{subequations}
With all these parametric relations in hand,we are left with 
\begin{equation}
\frac{1}{R_3^2}=\frac{b_0^2}{4},~~\frac{1}{R_1^2}=\frac{b_1^2}{4},~~\frac{1}{R_2^2}=\frac{b_2^2}{4},
\end{equation} which quite matches with the condition developed in 
from Einstein's equation by simplifying the stress-energy along with zero Ricci scalar.

Moreover, equation (\ref{triangle}) will thus generate a relation between the family of one parameter $\alpha$ and the flux parameters as
 \begin{equation}
     \alpha=\frac{b_1^2}{b_0^2},~~1-\alpha=\frac{b_2^2}{b_0^2}.
  \end{equation}
In the rest of the calculations, we will use this parameter $\alpha$ in deriving the string solutions and corresponding scaling relations. 
              
\subsection{Constraints and embeddings for rotating string ansatz}
We should have the embeddings for our background to satisfy the constraints for AdS and spherical geometries. From equation (\ref{triangle}) we can find the radii of the spheres as 
\begin{equation}
    R_1^2=\frac{R^2}{\alpha},~~R_2^{2}=\frac{R^2}{1-\alpha},~~\text{R=Radius of $AdS_3$}
\end{equation}Then the geometric constraints assume the forms
\begin{equation}
    -Y_{0}^{2}+Y_{1}^{2}+Y_{2}^{2}-Y_{3}^{2}=-R^2,~~X_{1}^{2}+X_{2}^{2}+X_{3}^{2}+X_{4}^{2}=\frac{R^2}{\alpha},~~W_{1}^{2}+W_{2}^{2}+W_{3}^{2}+W_{4}^{2}=\frac{R^2}{1-\alpha}
    \label{constraints}
\end{equation}respectively for $AdS_3$, $S^3_1$ and $S^3_2$. Here $Y_i$'s, $X_i$'s and $W_i$'s are the conventional notations for the embedding coordinates of $AdS_3$ and the two three spheres $S^3_1$ and $S^3_2$ respectively. The followings are the embeddings satisfying the above geometries
\begin{subequations}
\begin{align}
 &Y_{3}+iY_{0}=R\cosh{\rho}e^{it},~~Y_{1}+iY_{2}=R\sinh{\rho}e^{i\phi}\label{globalads} \ , \\&
    X_1+iX_2=\frac{R}{\sqrt{\alpha}}\cos{\beta_1}e^{i\beta_2},~~X_3+iX_4=\frac{R}{\sqrt{\alpha}}\sin{\beta_1}e^{i\beta_3}\label{global1} \ ,  \\&
   W_1+iW_2=\frac{R}{\sqrt{1-\alpha}}\cos{\beta_4}e^{i\beta_5},~~W_3+iW_4=\frac{R}{\sqrt{1-\alpha}}\sin{\beta_4}e^{i\beta_6}
   \label{global2}
\end{align}
\end{subequations}We shall cast our interest in the string dynamics along the coordinates of the spheres as well as those of $AdS_3$. Let us now define the embeddings $Y_i$'s, $X_i$'s and $W_i$'s in terms of the local coordinates as follows
\begin{subequations}
\begin{align}
&Y_3+iY_0=Rz_0(\xi)e^{ig_0(\xi)+i\omega_0\tau},~~Y_1+iY_2=Rz_1(\xi)e^{ig_1(\xi)+i\omega_1\tau}\label{localads} \ , \\&
   X_1+iX_2=\frac{R}{\sqrt{\alpha}}r_1(\xi)e^{if_1(\xi)+i\omega_2\tau},~~X_3+iX_4=\frac{R}{\sqrt{\alpha}}r_2(\xi)e^{if_2(\xi)+i\omega_3\tau}\label{locals1} \ , \\&
   W_1+iW_2=\frac{R}{\sqrt{1-\alpha}}r_3(\xi)e^{if_3(\xi)+i\omega_4\tau},~~W_3+iW_4=\frac{R}{\sqrt{1-\alpha}}r_4(\xi)e^{if_4(\xi)+i\omega_5\tau} \ .
   \label{locals2}
\end{align}
\end{subequations}
Here, the local coordinates $z_0, z_1$ and $r_i$'s with $i=1,2,3,4$ are the functions of the parameter $\xi=\alpha_1\sigma+\alpha_2\tau$, $\tau$ and $\sigma$ being the coordinates of the worldsheet swept out by the motion of the probe string under consideration while $\alpha_1$ and $\alpha_2$ are arbitrary constants. The constraints (\ref{constraints}) then yield 
\begin{equation}
    z_0^2-z_1^2=1
    \label{AdSconstraint}
\end{equation}for $AdS_3$ and
\begin{equation}
    \sum_{i=1,2}r_i^2=1,~~\sum_{j=3,4}r_j^2=1
\end{equation} for two spheres in the desired background. Comparing the relation (\ref{globalads}) with (\ref{localads}) we have, 
\begin{equation}
\cosh{\rho}=z_0,~~\sinh{\rho}=z_1,~~t=\phi_0,~~\phi=\phi_1;~~ \phi_i=g_i+\omega_i\tau, i=0,1.
\label{coordinates1}
\end{equation}  Similar comparison of relations (\ref{global1}) and (\ref{global2}) with (\ref{locals1}) and (\ref{locals2}) respectively achieve
\begin{equation}
\cos{\beta_1}=r_1,~~ \sin{\beta_1}=r_2,~~   \cos{\beta_4}=r_3,~~ \sin{\beta_4}=r_4
\label{coordinates2}
\end{equation}satisfying the constraints (\ref{constraints}) for the geometries of $S^3_1$ and $S^3_2$. Along with these we also get
\begin{equation}
    \phi_2=f_1+\omega_2\tau=\beta_2,~ \phi_3=f_2+\omega_3\tau=\beta_3,~ \phi_4=f_3+\omega_4\tau=\beta_5,~\phi_5=f_4+\omega_5\tau=\beta_6.
    \label{coordinates3}
\end{equation}
\section{Construction of Neumann-Rosochatius model}
 We wish to construct the one-dimensional NR integrable model for a probe fundamental string in the given $AdS_3\times S^3_1\times S^3_2\times S^1$ background in the presence of pure NSNS 2-form flux. To accomplish our aim, we will now explore a systematic formulation of different pieces, like Lagrangian, Hamiltonian and integrals of motion of our system by implementing the above embeddings.
    \subsection{Target space Lagrangian and Hamiltonian }
        To start with, let us consider the Polyakov action of fundamental string with WZ interaction term given as
    \begin{eqnarray}
        S=-\frac{T}{2}\int d\tau d\sigma \sqrt{-\gamma}
        \gamma^{\alpha\beta}G_{MN}\partial_{\alpha}X^M\partial_{\beta}X^N+\frac{T}{2}
        \int d\sigma d\tau \epsilon^{\alpha\beta}B_{MN}\partial_\alpha X^M\partial_\beta X^N \ , \nonumber \\
    \label{Polyakov}
    \end{eqnarray}
where $T$ is the string tension, $\gamma_{\alpha\beta}$ is the auxiliary metric which represents the intrinsic geometry of the background and can be written in conformal gauge as $\gamma_{\alpha\beta}=\eta_{\alpha\beta}$=diag(-1,+1) and $\epsilon^{\tau \sigma}=-\epsilon^{\sigma \tau}=1$. The 10-dimensional background coordinates are denoted by the indices $M$, $N$ and the 2-dimensional worldsheet coordinates are denoted by the indices $\alpha$, $\beta$. $G_{MN}$ and $B_{MN}$ represent the induced metric of the background and the anti-symmetric pullback of the finite flux respectively.
\begin{equation}
    B_{t\phi}=b_0R^2z_1^2,~~
    B_{\beta_2\beta_3}=-b_1\frac{R^2}{\alpha}r_2^2,~~ 
    B_{\beta_5\beta_6}=-b_2\frac{R^2}{(1-\alpha)}r_4^2.
   \end{equation}
   The target space Lagragian obtained from the Polyakov action may be written as (\ref{Polyakov})
    \begin{equation}
        \begin{split}
    \mathcal{L}=&-\frac{T}{2}\Big[-\partial_{a}Y_0\partial^aY_0^*+\partial_{a}Y_1\partial^aY_1^*+\partial_{a}X_1\partial^aX_1^*+\partial_{a}X_2\partial^aX_2^*+\partial_{a}W_1\partial^aW_1^*\\&+\partial_{a}W_2\partial^aW_2^*\Big]-\frac{\Lambda}{2}\left(|Y_0|^2-|Y_1|^2-1\right)-\frac{\tilde\Lambda_1}{2}\left(|X_1|^2+|X_2|^2-1\right)\\&-\frac{\tilde\Lambda_2}{2}\left(|W_1|^2+|W_2|^2-1\right)+\frac{T}{2}\epsilon^{ab}B_{MN}\partial_aX^M\partial_bX^N.
    \label{Lagrangian}
    \end{split}
    \end{equation}
Using equations (\ref{coordinates1}), (\ref{coordinates2}) and (\ref{coordinates3}) and subsituting for the derivatives of the coordinates in the Lagrangian (\ref{Lagrangian}), we get
\begin{equation}
    \begin{split}
        &\mathcal{L}=-\frac{TR^2}{2}\Big[(\alpha_2^2-\alpha_1^2)(z_0^{\prime 2}-z_1^{\prime 2})+z_0^2(\alpha_2^2-\alpha_1^2)\left(g_0^{\prime}+\frac{\alpha_2 \omega_0}{\alpha_2^2-\alpha_1^2}\right)^2-\frac{\alpha_1^2z_0^2\omega_0^2}{\alpha_2^2-\alpha_1^2}+\frac{\alpha_1^2z_1^2\omega_1^2}{\alpha_2^2-\alpha_1^2}\\& -z_1^2(\alpha_2^2-\alpha_1^2)\left(g_1^{\prime}+\frac{\alpha_2 \omega_1}{\alpha_2^2-\alpha_1^2}\right)^2\Big]-\frac{\tilde{\Lambda}}{2}\left(g^{ab}z_az_b+1\right)-\frac{TR^2}{2\alpha}\Big[(\alpha_1^2-\alpha_2^2)(r_1^{\prime 2}+r_2^{\prime 2})\\&+\sum_{i=1}^2 r_i^2(\alpha_1^2-\alpha_2^2)\left(f_i^{\prime} -\frac{\alpha_2\omega_{i+1}}{(\alpha_1^2-\alpha_2^2)}\right)^2-\sum_{i=1}^2\frac{\alpha_1^2r_i^2\omega_{i+1}^2}{(\alpha_1^2-\alpha_2^2)}\Big]-\frac{\Lambda_1}{2}(r_1^2+r_2^2-1) \\&-\frac{TR^2}{2(1-\alpha)}\Big[(\alpha_1^2-\alpha_2^2)(r_3^{\prime 2}+r_4^{\prime 2})+\sum_{p=3}^4 r_p^2(\alpha_1^2-\alpha_2^2)\left(f_p^{\prime} -\frac{\alpha_2\omega_{p+1}}{(\alpha_1^2-\alpha_2^2)}\right)^2-\sum_{p=1}^2\frac{\alpha_1^2r_p^2\omega_{p+1}^2}{(\alpha_1^2-\alpha_2^2)}\Big]\\&-\frac{\Lambda_2}{2}(r_3^2+r_4^2-1) +TR^2\alpha_1\bigg[-b_1\frac{r_2^2}{\alpha}\left(\omega_2f_2^{'}-\omega_3f_1^{'}\right)-b_2\frac{r_4^2}{(1-\alpha)}\left(\omega_4f_4^{'}-\omega_5f_3^{'}\right) \\& +b_0z_1^2\left(\omega_0g_1^{'}-\omega_{1}g_0^{'}\right)\bigg].
        \label{targetspace}
    \end{split}
\end{equation}

We will now derive the equations of motion corresponding to various local coordinates present in the target-space Lagragian (\ref{targetspace}) by means of Euler-Lagrange equation.\\

\textbf{Equations of motion for the embeddings for $AdS_3$ :}\\

The equations of motion for $z_a$ for $a=0,1$ can be derived as
\begin{equation}
    \left(\alpha_1^2-\alpha_2^2\right)z_0^{''}=\left[\left(\alpha_1^2-\alpha_2^2\right)g_0^{'2}-2\alpha_2\omega_0g_0^{'}-\omega_0^2+\frac{\tilde{\Lambda}}{TR^2}\right]z_0,
    \label{adsequation1}
\end{equation}and
\begin{equation}
        \left(\alpha_1^2-\alpha_2^2\right)z_1^{''}=\left[\left(\alpha_1^2-\alpha_2^2\right)g_1^{'2}-2\alpha_2\omega_1g_1^{'}-\omega_1^2+2\alpha_1b_0\left(\omega_0g_1^{'}-\omega_{1}g_0^{'}\right)-\frac{\tilde{\Lambda}}{TR^2}\right]z_1,
        \label{adsequation2}
\end{equation}
where the angular coordinates $g_a$'s satisfy the relations
\begin{equation}
g_0^{'}=\frac{1}{\alpha_1^2-\alpha_2^2}\left[\frac{C_0}{z_0^2}+\alpha_2\omega_0+\alpha_1\omega_1b_0\frac{z_1^2}{z_0^2}\right],
\label{adsequation3}
\end{equation}and
\begin{equation}
g_1^{'}=\frac{1}{\alpha_1^2-\alpha_2^2}\left[\alpha_2\omega_1+\alpha_1b_0\omega_0-\frac{C_1}{z_1^2}\right], 
\label{adsequation4}
\end{equation}
$C_0$ and $C_1$ being the integration constants. Substituting the expressions of $g_0^{'}$ and $g_1^{'}$ in the Lagragian we get 
\begin{equation}
\begin{split}
    &\mathcal{L}_{AdS_3}=-\frac{TR^2}{2}\left(\alpha_1^2-\alpha_2^2\right)\Bigg[-z_0^{'2}+z_1^{'2}-\frac{\left(C_0+\alpha_1\omega_1b_0 z_1^2\right)^2}{\left(\alpha_1^2-\alpha_2^2\right)^2 z_0^{2}}+\frac{\left(\alpha_1\omega_0b_0 z_1^2-C_1\right)^2}{\left(\alpha_1^2-\alpha_2^2\right)^2 z_1^{2}}\\&+\frac{\alpha_1^2}{\left(\alpha_1^2-\alpha_2^2\right)^2}{\left(\omega_0^2z_0^2-\omega_1^2z_1^2\right)}\Bigg].
    +\frac{TR^2\alpha_1b_0z_1^2}{\left(\alpha_1^2-\alpha_2^2\right)^2}\Big(\alpha_1b_0\omega_0^2-\frac{\omega_0 C_1}{z_1^2}-\frac{\omega_1C_0}{z_0^2}-\frac{\alpha_1b_0\omega_1^2z_1^2}{z_0^2}\Big)\\&-\frac{\tilde{\Lambda}}{2}\left(g^{ab}z_az_b+1\right)
    \end{split}
    \label{AdSL}
\end{equation}
\textbf{Equations of motion for the embeddings for $S^3_1$ :}\\

The equations of motion for $r_i$ for $i=1,2$ can be derived as
 \begin{equation}
     r_1^{''}=\left[f_1^{'2}-\frac{\left(\omega_2^2+2\alpha_2\omega_2f_1^{'}\right)}{(\alpha_1^2-\alpha_2^2)}+\frac{\alpha\Lambda_1}{TR^2(\alpha_1^2-\alpha_2^2)}\right]r_1,
 \end{equation}
 \begin{equation}
      r_2^{''}=\left[f_2^{'2}-\frac{\left(\omega_3^2+2\alpha_2\omega_3f_2^{'}\right)}{(\alpha_1^2-\alpha_2^2)}+\frac{2\alpha_1b_1}{(\alpha_1^2-\alpha_2^2)}\left(\omega_2f_2^{'}-\omega_3f_1^{'}\right)+\frac{\alpha\Lambda_1}{TR^2\left(\alpha_1^2-\alpha_2^2\right)}\right]r_2.
 \end{equation} Where the angular coordinates $f_i$'s satisfy the relations
 \begin{equation}
     f_1^{'}=\frac{1}{(\alpha_1^2-\alpha_2^2)}\left[\frac{v_1}{r_1^2}+\alpha_2\omega_2+\frac{\alpha_1b_1\omega_3r_2^2}{r_1^2}\right]
 \end{equation}and
 \begin{equation}
     f_2^{'}=\frac{1}{(\alpha_1^2-\alpha_2^2)}\left[\frac{v_2}{r_2^2}+\alpha_2\omega_3-\alpha_1b_1\omega_2\right]
 \end{equation}Here $v_i$'s are suitable constants for integration. Substituting $f_1^{'}$ and $f_2^{'}$ will give us the Lagrangian for $S^3_1$ as
 \begin{equation}
 \begin{split}
 \mathcal{L}_{S^3_1}=&- \frac{TR^2}{2\alpha}(\alpha_1^2-\alpha_2^2)\sum_{i=1}^{2}\left[r_i^{'2}+\frac{r_i^{-2}}{(\alpha_1^2-\alpha_2^2)^2}\left(v_i+b_1\alpha_1 r_2^2\epsilon^{ij}\omega_{j+1}\right)^2-\frac{\alpha_1^2\omega^2_{i+1}r_i^2}{(\alpha_1^2-\alpha_2^2)^2}\right]\\&+\frac{TR^2\alpha_1 r_2^2b_1}{\alpha (\alpha_1^2-\alpha_2^2)}\sum_{i=1}^{2}\left(\epsilon^{ij}\frac{v_i\omega_{j+1}}{r_i^2}+\frac{\alpha_1 b_1 \omega_{3}^2r_2^2}{r_1^2}+b_1\alpha_1 \omega_2^2\right)-\frac{\Lambda_1}{2}\left(\sum_{i=1}^{2}r_i^2-1\right)
 \label{S1L}
 \end{split}
 \end{equation}
\textbf{Equations of motion for the embeddings for $S^3_2$ :}\\

The equations of motion for $r_p$ for $p=3,4$ can be derived as
 \begin{equation}
     r_3^{''}=\left[f_3^{'2}-\frac{\left(\omega_4^2+2\alpha_2\omega_4f_3^{'}\right)}{(\alpha_1^2-\alpha_2^2)}+\frac{(1-\alpha)\Lambda_2}{TR^2(\alpha_1^2-\alpha_2^2)}\right]r_3
 \end{equation}and
 \begin{equation}
      r_4^{''}=\left[f_4^{'2}-\frac{\left(\omega_5^2+2\alpha_2\omega_5f_4^{'}\right)}{(\alpha_1^2-\alpha_2^2)}+\frac{2\alpha_1b_2}{(\alpha_1^2-\alpha_2^2)}\left(\omega_4f_4^{'}-\omega_5f_3^{'}\right)+\frac{(1-\alpha)\Lambda_2}{TR^2(\alpha_1^2-\alpha_2^2)}\right]r_4 
 \end{equation}whereas the angular coordinates $f_i$'s satisfy the relations
 \begin{equation}
     f_3^{'}=\frac{1}{(\alpha_1^2-\alpha_2^2)}\left[\frac{v_3}{r_3^2}+\alpha_2\omega_4+\frac{\alpha_1b_2\omega_5r_4^2}{r_3^2}\right]
 \end{equation}and
 \begin{equation}
     f_4^{'}=\frac{1}{(\alpha_1^2-\alpha_2^2)}\left[\frac{v_4}{r_4^2}+\alpha_2\omega_5-\alpha_1b_2\omega_4\right]
 \end{equation} The Lagrangian for $S^3_2$ is obtained with these expressions of $f_p$'s as,
 \begin{equation}
 \begin{split}
 \mathcal{L}_{S^3_2}&=- \frac{TR^2}{2(1-\alpha)}(\alpha_1^2-\alpha_2^2)\sum_{p=3}^{4}\left[r_p^{'2}+\frac{r_p^{-2}}{(\alpha_1^2-\alpha_2^2)^2}\left(v_p+b_2\alpha_1 r_4^2\epsilon^{pq}\omega_{q+1}\right)^2-\frac{\alpha_1^2\omega^2_{p+1}r_p^2}{(\alpha_1^2-\alpha_2^2)^2}\right]\\&+\frac{TR^2b_2\alpha_1 r_4^2}{(1-\alpha)(\alpha_1^2-\alpha_2^2)}\sum_{p=3}^{4}\left(\epsilon^{pq}\frac{v_p\omega_{q+1}}{r_p^2}+\frac{b_2\alpha_1 \omega_{5}^2r_4^2}{r_3^2}+b
 _2\alpha_1 \omega_4^2\right)-\frac{\Lambda_2}{2}\left(\sum_{p=3}^{4}r_p^2-1\right)
 \end{split}
 \label{S2L}
 \end{equation}
 The total Lagrangian is (\ref{AdSL}), (\ref{S1L}) and (\ref{S2L}) as
 \begin{equation}
    \mathcal{L}= \mathcal{L}_{AdS_3}+\mathcal{L}_{S^3_1}+\mathcal{L}_{S^3_2}.
 \end{equation}
 Here it is to be noted that all the finite fluxes turned over on different geometries of the background are incorporated in the corresponding Lagrangians in those geometries. Evidently, the terms along with those due to fluxes in the Lagragian involves either $x^2$-type harmonic oscillator potential or $\frac{1}{x^2}$-type centrifugal potential barrier, $x$ denoting the position coordinates in respective geometries. Hence, it is obvious that the total Lagrangian of a probe fundamental string in $AdS_3\times S^3_1\times S^3_2$ background having pure NSNS 2-form flux quite resembles with the Lagrangian of one-dimensional integrable NR model, except having some integrable deformations added to it due to the finite flux. 

\subsection{Integrals of motion}
One of the necessary conditions that an integrable system should consists of is the presence of infinite tower of conserved quantities in involution, namely the integrals of motion. K. Uhlenbeck first put forth the generalised form of the integrals of motion along with some deformations for the one-dimensional classical NR integrable system containing N number of harmonic oscillators and these are expressed as
\begin{equation}
I_{i}=\alpha_1^2 x_{i}^{2}+\sum_{j\neq i}\frac{1}{\omega_{i}^{2}-\omega_{j}^{2}}\left[\left(x_{i}x_{j}^{'}-x_{j}x_{i}^{'}\right)^{2}+\pi_{\alpha_ i}^{2}\frac{x_{j}^{2}}{x_{i}^{2}}+\pi_{\alpha_ j}^{2}\frac{x_{i}^{2}}{x_{j}^{2}}\right], \
\label{Integrals}
\end{equation}which satisfy the relations $\left\{I_{i},I_{j}\right\} =0 ~  \forall  ~i,j \in \left\{1,2,......,N\right\}$ and $\sum_{i=1}^{N}I_i=\alpha_1^2 $. In our study we will also check the existence of such integrals of motion for our system to validate the construction of integrable NR model. It is to be noted that the Uhlenbeck integrals of motion would contain some deformations caused by the finite fluxes in different geometry of our background. But these deformations should be in such a way that the deformed integrals of motion should be conserved quantities following the necessary conditions. To calculate the deformation we write the extended expression of the integrals of motion as
    \begin{equation}
        \bar{I}_i= \alpha_1^2 x_{i}^{2}+\sum_{j\neq i}\frac{1}{\omega_{i}^{2}-\omega_{j}^{2}}\left[\left(x_{i}x_{j}^{'}-x_{j}x_{i}^{'}\right)^{2}+\pi_{\alpha_ i}^{2}\frac{x_{j}^{2}}{x_{i}^{2}}+\pi_{\alpha_ j}^{2}\frac{x_{i}^{2}}{x_{j}^{2}}+2f\right],
         \label{IOM2}
    \end{equation}
where $x_i$'s takes the form $x_i(\xi)=r_i(\xi)e^{if_i(\xi)+i\omega_i\tau}$ in polar coordinates, $\pi_i$'s are corresponding canonically conjugate momenta and $f$ is the deformation which is independent of any terms like $x_i^ \prime$ but depends on the flux parameter $b_0,b_1,b_2$. We can calculate this function by imposing the condition that $\bar I_i^\prime =0$. After some quick algebra along with using the constraints and some equations of motions, we find the the function to appear as total derivatives, integrating which we can immediately get the deformed Uhlenbeck constant. Like the constants in the absence of the flux, the deformed constants also satisfy the condition
$\sum_{i=1}^N \bar I_i=\alpha_1^2(1-q^2)$.

Let us start with inserting the expression for $x_i$ and $p_i$ in equation (\ref{IOM2}) and finding the integrals of motion for $AdS_3$. In this case we have two integrals $\bar I_1$ and $\bar I_2,$ but due to the constraint they satisfy i.e, $\bar I_1- \bar I_0=-\alpha_1^2(1-q^2)$ (for $AdS_3$), we are left with a single independent constant,
\begin{equation}
    \bar I_1=\alpha_1^2 z_0^2+\frac{1}{\omega_1^2-\omega_0^2}\left[(\alpha_1^2 - \alpha_2^2)^2 (z_1z_0^{'}-z_0z_1^{'})^2+\frac{C_0^2z_1^2}{z_0^2}+\frac{C_1^2z_0^2}{z_1^2}+2f\right]
\end{equation}
We will now determine $f$ by imposing $\bar{I}_1^{'}=0$ and using the constraint $z_1^2-z_0^2=-1$ together with the equation (\ref{adsequation1}),(\ref{adsequation2}),(\ref{adsequation3}) and (\ref{adsequation4}) as follows
\begin{equation}
    f^{'}+\alpha_1^2b_0^2(\omega_1^2-\omega_0^2)z_0z_0^{'}+\frac{z_0^{'}\left(2\alpha_1b_0\omega_1C_0-\alpha_1^2b_0^2\omega_1^2\right)}{z_0^3}=0
\end{equation}
We can see from the above expression that the differential equation in $f$ is a total derivative which can be immediately integrated to give the form of the deformed Uhlenbeck integrals of motion as
\begin{equation}
\begin{split}
    \bar I_1=\alpha_1^2~(1-b_0^2)~ z_0^2+\frac{1}{\omega_1^2-\omega_0^2}&\Bigg[(\alpha_1^2 - \alpha_2^2)^2 (z_1z_0^{'}-z_0z_1^{'})^2+\frac{(C_0-\alpha_1b_0\omega_1)^2z_1^2}{z_0^2}+\frac{C_1^2z_0^2}{z_1^2}\Bigg],
    \end{split}
\end{equation}
where we have neglected the constant term appearing due to the presence of flux, in order to write the expression in terms of complete square. Proceeding in the similar fashion for the two spheres, 
we find the deformed Uhlenbeck integrals of motion satisfying the constraint $\bar I_2+\bar I_3=\alpha_1^2(1-q^2)$ and $\bar I_3+\bar I_4=\alpha_1^2(1-q^2)$ respectively.
\begin{equation}
\begin{split}
    \bar I_2=\alpha_1^2~(1-b_1^2)~ r_1^2+\frac{1}{\omega_2^2-\omega_3^2}&\Bigg[(\alpha_1^2 - \alpha_2^2)^2 (r_1r_2^{'}-r_2r_1^{'})^2+\frac{(v_1+\alpha_1b_1\omega_3)^2r_2^2}{r_1^2}+\frac{v_2^2r_1^2}{r_2^2}\Bigg],
    \end{split}
\end{equation}
and
\begin{equation}
\begin{split}
    \bar I_4=\alpha_1^2~(1-b_2^2)~ r_3^2+\frac{1}{\omega_4^2-\omega_5^2}&\Bigg[(\alpha_1^2 - \alpha_2^2)^2 (r_3r_4^{'}-r_4r_3^{'})^2+\frac{(v_3^2+\alpha_1b_2\omega_5)^2r_4^2}{(r_3^2}+\frac{v_4^2r_3^2}{r_4^2}\Bigg].
    \end{split}
\end{equation}
    We can see in the above expressions for the deformed integrals of motions that they satisfy the constraint i.e, $\sum_i I_i=\alpha_1^2$, only for pure RR case. It may also be noted that the Hamiltonian associated with the NR model can be expressed as the linear combination of integrals of motion of the system. The background in our case have three different independent integrals of motions. To be precise, for each geometry i.e., $AdS_3, S_3^1,S_3^2$, we have two coordinates, two phases, their corresponding conjugate momenta and two constraints. So in order to solve the system under consideration, classical Liouville integrability demands 3 Uhlenbeck constants to be in involution. Among these the canonically conjugate momenta of two cyclic phases in the Lagrangian naturally gives two conserved quantities. The remaining 2 are the deformed Uhlenbeck integrals of motion found above. But the two Uhlenback integrals of motion satisfy the following conditions, $\bar I_1- \bar I_0=-\alpha_1^2 (1-q^2)$ for $AdS_3$ and $\sum \bar I_i=\alpha_1^2 (1-q^2)$ for the two spheres. This clearly tells that one of the two quantities is independent for each geometry. 
 \section{Rotating string solutions}
   In this section we will consider the spiky string in $S_1^3$ with non-zero angular momentum $J$ and winding $m$ in $S^1 \subset S_2^3$. We start with the Lagrangian for the whole background and then consider the part which deals with the dynamics of the string. Here the differentiation is with respect to $\xi$ which is a function of $\sigma,\tau$ in the form $\xi=\alpha_1\sigma+\alpha_2\tau$. In the embedding coordinates, the Lagrangian takes the following form:
    \begin{equation}
        \begin{split}
    \mathcal{L}=&-\frac{T}{2}\Big[-\partial_{a}Y_0\partial^aY_0^*+\partial_{a}Y_1\partial^aY_1^*+\partial_{a}X_1\partial^aX_1^*+\partial_{a}X_2\partial^aX_2^*+\partial_{a}W_1\partial^aW_1^*\\&+\partial_{a}W_2\partial^aW_2^*+\Lambda\left(|Y_0|^2-|Y_1|^2-1\right)+\tilde\Lambda_1\left(|X_1|^2-|X_2|^2-1\right)\\&+\tilde\Lambda_2\left(|W_1|^2-|W_2|^2-1\right)\Big]+\frac{T}{2}\epsilon^{ab}B_{MN}\partial_aX^M\partial_bX^N
    \end{split}
    \end{equation}
     Using (\ref{localads},\ref{locals1},\ref{locals2}) the lagrangian can be written as
        \begin{equation}
    \begin{split}
        \mathcal{L}&=-\frac{TR^2}{2}\Big[(\alpha_2^2-\alpha_1^2)(z_0^{\prime 2}-z_1^{\prime 2})+z_0^2(\alpha_2^2-\alpha_1^2)\left(g_0^{\prime}+\frac{\alpha_2 \omega_0}{\alpha_2^2-\alpha_1^2}\right)^2-\frac{\alpha_1^2z_0^2\omega_0^2}{\alpha_2^2-\alpha_1^2}+\frac{\alpha_1^2z_1^2\omega_1^2}{\alpha_2^2-\alpha_1^2}\\&-z_1^2(\alpha_2^2-\alpha_1^2)\left(g_1^{\prime}+\frac{\alpha_2 \omega_1}{\alpha_2^2-\alpha_1^2}\right)^2+\frac{\tilde{\Lambda}}{2}\left(g^{ab}z_az_b+1\right)\Big]-\frac{TR^2}{2\alpha}\Big[(\alpha_1^2-\alpha_2^2)(r_1^{\prime 2}+r_2^{\prime 2})\\&+\sum_{i=1}^2 r_i^2(\alpha_1^2-\alpha_2^2)\left(f_i^{\prime} -\frac{\alpha_2\omega_{i+1}}{(\alpha_1^2-\alpha_2^2)}\right)^2-\sum_{i=1}^2\frac{\alpha_1^2r_i^2\omega_{i+1}^2}{(\alpha_1^2-\alpha_2^2)}+\frac{\Lambda_1}{2}(r_1^2+r_2^2-1) \Big]\\&-\frac{TR^2}{2(1-\alpha)}\Big[(\alpha_1^2-\alpha_2^2)(r_3^{\prime 2}+r_4^{\prime 2})+\sum_{p=3}^4 r_p^2(\alpha_1^2-\alpha_2^2)\left(f_p^{\prime} -\frac{\alpha_2\omega_{p+1}}{(\alpha_1^2-\alpha_2^2)}\right)^2-\sum_{p=1}^2\frac{\alpha_1^2r_p^2\omega_{p+1}^2}{(\alpha_1^2-\alpha_2^2)}\\&+\frac{\Lambda_1}{2}(r_3^2+r_4^2-1) \Big]+TR^2\alpha_1\Big[-b_1\frac{r_2^2}{\alpha}\left(\omega_2f_2^{'}-\omega_3f_1^{'}\right)-b_2\frac{r_4^2}{1-\alpha}\left(\omega_4f_4^{'}-\omega_5f_3^{'}\right) \\& +b_0z_1^2\left(\omega_0g_1^{'}-\omega_{1}g_0^{'}\right)\Big].
    \end{split}
\end{equation}
The conformal constraints are
    \begin{equation*}
        -|\dot{Y_0}|^2+|\dot Y_1|^2+|\dot X_i|^2+|\dot W_i|^2-|{Y_0}^\prime|^2+|Y_1^\prime|^2+|X_i^\prime|^2+|W_i^\prime|^2=0,~~~~~~~~i=1,2 
    \end{equation*}
    and
    \begin{equation}
        -\dot Y_0 Y_0^{\prime*}+\dot Y_1 Y_1^{\prime*}+\dot X_i X_i^{\prime*}+\dot W_i W_i^{\prime*}+c.c=0.
        \end{equation}
    Now we want the string profile for the spiky string spinning in $S_1^3$ along with the additional non-zero angular momentum $J$ in $S^1\subset S_2^3$. This can be done by taking $Y_1$ and $W_i$ to be constant. The radius of the two spheres are related by a parameter $\alpha$ as discussed earlier. Note that the radius of $AdS_3, S_1$ and $S_2$ are $R,\frac{R}{\sqrt{\alpha}}$ and $\frac{R}{\sqrt{1-\alpha}}$ respectively. The metric of the system under consideration is then given by
    \begin{equation}
            ds^2=-dY_0d\bar{Y}_0+\sum_{i=1}^2 dX_i d\bar{X}_i+dZd\bar{Z}.
        \end{equation}    
    We consider the following NR ansatz
        \begin{equation}
        \begin{split}
           Y_0+&iY_3=Re^{i\omega_0\tau},~~~~X_1+iX_2=\frac{R}{\sqrt{\alpha}}r_1(\xi)e^{i(f_1(\xi)+\omega_2\tau)},\\&X_3+iX_4=\frac{R}{\sqrt{\alpha}}r_2(\xi)e^{i(f_2(\xi)+\omega_3\tau)},~~~~Z_1+iZ_2=\frac{R}{\sqrt{1-\alpha}}e^{i(\psi(\xi)+\nu\tau)}.
           \end{split}
        \end{equation}
        With this ansatz, the string Lagrangian reduces to the following 1-dimensional integrable system,
        \begin{equation}
    \begin{split}
        \mathcal{L}=&-\frac{TR^2}{2}\omega_0^2-\frac{TR^2}{2\alpha}\Bigg[(\alpha_1^2-\alpha_2^2)(r_1^{\prime 2}+r_2^{\prime 2})+\sum_{i=1}^2 r_i^2(\alpha_1^2-\alpha_2^2)\left(f_i^{\prime} -\frac{\alpha_2\omega_{i+1}}{(\alpha_1^2-\alpha_2^2)}\right)^2\\&-\sum_{i=1}^2\frac{\alpha_1^2r_i^2\omega_{i+1}^2}{(\alpha_1^2-\alpha_2^2)}+\frac{\Lambda_1}{2}(r_1^2+r_2^2-1) \Bigg]-\frac{TR^2}{2(1-\alpha)}\Bigg[-\frac{\alpha_1^2\nu^2}{(\alpha_1^2-\alpha_2^2)}\\&+(\alpha_1^2-\alpha_2^2)\left(\psi^{\prime} -\frac{\alpha_2\nu}{(\alpha_1^2-\alpha_2^2)}\right)^2 \Bigg].
        \end{split}
   \end{equation}  
    Here we have considered pure RR case. Also it can be seen from the ansatz that, 
    $t=\omega_0 \tau$. We impose the periodicity conditions on the parameters to find solutions of the closed string as 
    \begin{equation}
    r_i(\xi)=r_i(\xi+2\pi\alpha_1),\psi(\xi)=\psi(\xi+2\pi\alpha_1)-2\pi m, f_i(\xi)=f_i(\xi+2\pi\alpha_1)-2\pi m_i,~~ i=1,2.
    \end{equation}
    where $m,m_1$ and $m_2$ are taken to be integers.
    The Euler-Lagrange equations of motion for $f_i$ and $\psi$ are given by
    \begin{equation}
         f_1^{'}=\frac{1}{(\alpha_1^2-\alpha_2^2)}\left[\frac{v_1}{r_1^2}+\alpha_2\omega_2\right]
,~
     f_2^{'}=\frac{1}{(\alpha_1^2-\alpha_2^2)}\left[\frac{v_2}{r_2^2}+\alpha_2\omega_3\right],~ \psi^{'}=\frac{v+\alpha_2\nu}{\alpha_1^2-\alpha_2^2}
    \end{equation}
    where $v_1,v_2,v$ are the constants of integration. Let us write the angle in $S^1\subset S^3$ as $\theta$, so we can write
       \begin{equation}
           Z=e^{i\theta}, ~~~~~~\theta=\psi+\nu \tau=\frac{v+\alpha_2\nu}{\alpha_1^2-\alpha_2^2}\xi.
       \end{equation}
       Now we define the winding number in $\theta$ to be of the form
       \begin{equation}
           2\pi m=\int_{0}^{2 \pi \alpha_1}d\xi \psi{'}= \frac{v+\alpha_2\nu}{\alpha_1^2-\alpha_2^2}\int d\xi.
           \label{winding}
       \end{equation}
       We do not take any winding in $\phi_3$ direction. This gives the following condition
       \begin{equation}
          2\pi m_2= \int_0^{2\pi \alpha_1}d\xi f_2^{'}=\int_0^{2\pi\alpha_1}d\xi\left(\frac{v_2}{r_2^2}+\alpha_2\omega_3\right)=0.
          \label{zerowinding}
       \end{equation}
       The expression for lagrangian with parameters $r_1,r_2$ can be expressed as
       \begin{equation}
    \begin{split}
        L=-&\frac{TR^2}{2}\omega_0^2-\frac{TR^2}{2\alpha}\Bigg[(\alpha_1^2-\alpha_2^2)(r_1^{\prime 2}+r_2^{\prime 2})+ \frac{r_1^{-2}v_1^2}{\alpha_1^2-\alpha_2^2}-\sum_{i=1}^2\frac{\alpha_1^2r_i^2\omega_{i+1}^2}{(\alpha_1^2-\alpha_2^2)}\\&+\frac{r_2^{-2}v_2^2}{\alpha_1^2-\alpha_2^2}+\frac{\Lambda_1}{2}(r_1^2+r_2^2-1) \Bigg]-\frac{TR^2}{2(1-\alpha)}\Bigg[\frac{v^2}{\alpha_1^2-\alpha_2^2}-\frac{\alpha_1^2\nu^2}{\alpha_1^2-\alpha_2^2} \Bigg].
        \end{split}
        \end{equation}
        The corresponding Hamiltonian then becomes
         \begin{equation}
    \begin{split}
        H=-\frac{TR^2}{2\alpha}&\Bigg[(\alpha_1^2-\alpha_2^2)(r_1^{\prime 2}+r_2^{\prime 2})- \frac{r_1^{-2}v_1^2}{\alpha_1^2-\alpha_2^2}+\sum_{i=1}^2\frac{\alpha_1^2r_i^2\omega_{i+1}^2}{(\alpha_1^2-\alpha_2^2)}-\frac{r_2^{-2}v_2^2}{\alpha_1^2-\alpha_2^2}\Bigg]\\&-\frac{TR^2}{2(1-\alpha)}\Bigg[-\frac{v^2}{\alpha_1^2-\alpha_2^2}+\frac{\alpha_1^2\nu^2}{\alpha_1^2-\alpha_2^2} \Bigg].
        \label{Hamiltonian}
        \end{split}
        \end{equation}
        Calculating the first and second Virasoro constraints we get,
        \begin{equation}
        \begin{split}
            (\alpha_1^2+\alpha_2^2)&\left(r_1^{\prime 2}+r_2^{\prime 2}\right)+(\alpha_1^2+\alpha_2^2)\left(r_1^2f_1^{\prime 2}+r_2^2f_2^{\prime 2}+\psi^{\prime 2}\frac{\alpha}{1-\alpha}\right)+\omega_{2}^2r_1^2+\omega_{3}^2r_2^2\\&+\nu^2\frac{\alpha}{1-\alpha}+2\alpha_2\left(\omega_2f_1^\prime r_1^2+\omega_3f_2^\prime r_2^2+\nu \psi^ \prime\frac{\alpha}{1-\alpha}\right)=\omega_0^2\alpha,
            \label{firstvirasoro}
        \end{split}
        \end{equation}
        and
        \begin{equation}
            \begin{split}
                \alpha_2\left(r_1^{\prime 2}+r_2^{\prime 2}+r_1^2f_1^{\prime 2}+r_2^2f_2^{\prime 2}+\frac{\psi^{\prime 2}\alpha}{1-\alpha}\right)=-\left(\omega_2f_1^\prime r_1^2+\omega_3f_2^\prime r_2^2+\nu \psi^{\prime}\frac{\alpha}{1-\alpha}\right).
                \label{secondvirasoro}
            \end{split}
        \end{equation}
        Simplifying (\ref{secondvirasoro}) and substituting in (\ref{firstvirasoro}) we get,
           \begin{equation}
        \begin{split}
            (\alpha_1^2-\alpha_2^2)\left(r_1^{\prime 2}+r_2^{\prime 2}+r_1^2f_1^{\prime 2}+r_2^2f_2^{\prime 2}+\frac{\psi^{\prime 2}\alpha}{1-\alpha}\right)+\omega_{2}^2r_1^2+\omega_{3}^2r_2^2+\frac{\nu^2\alpha}{1-\alpha}=\omega_0^2\alpha
            \label{virasoro}
        \end{split}
        \end{equation}
        and
        \begin{equation}
           \omega_2v_1+\omega_3v_2+\nu v\frac{\alpha}{1-\alpha}=-\alpha_2\omega_0^2\alpha.
           \label{secondconstraint}
        \end{equation}
    The first constraint leads to the form of the Hamiltonian to be
    \begin{equation}
        H=\omega_0^2\alpha\Big(\frac{\alpha_1^2+\alpha_2^2}{\alpha_1^2-\alpha_2^2}\Big).
        \label{Constrainthamiltonian}
    \end{equation}
    The conserved charges can be written as
        \begin{equation}
                  E=TR^2\int \frac{\omega_0}{\alpha_1}d\xi~,~~J=\frac{TR^2}{1-\alpha}\int \frac{d\xi}{\alpha_1}\left(\frac{v\alpha_2}{\alpha_1^2-\alpha_2^2}+\frac{\alpha_1^2\nu}{\alpha_1^2-\alpha_2^2}\right),
        \end{equation}
        \begin{equation}
            J_1=\frac{TR^2}{\alpha}\int\frac{d\xi}{\alpha_1}\Bigg(\frac{v_1\alpha_2}{\alpha_1^2-\alpha_2^2}+\frac{\omega_2\alpha_1^2r_1^2}{\alpha_1^2-\alpha_2^2}\Bigg),
            J_2=\frac{TR^2}{\alpha}\int\frac{d\xi}{\alpha_1}\Bigg(\frac{v_2\alpha_2}{\alpha_1^2-\alpha_2^2}+\frac{\omega_3\alpha_1^2r_2^2}{\alpha_1^2-\alpha_2^2}\Bigg),
        \end{equation}
        Now we will use the condition, $H$ being conserved along with the constraint (\ref{Constrainthamiltonian}) and the relation $r_1^2+r_2^2=1$ in (\ref{Hamiltonian}), we get 
        \begin{equation}
        \begin{split}
           \omega_0^2\alpha\left(\frac{1+\alpha_2^2}{1-\alpha_2^2}\right)=(1-\alpha_2^2)&\frac{r_2^{\prime 2}}{1-r_2^2}-\frac{1}{1-\alpha_2^2}\left(\frac{v_1^2}{1-r_2^2}+\frac{v_2^2}{r_2^2}+\frac{\alpha v^2}{1-\alpha}\right)\\&+\frac{1}{1-\alpha_2^2}\left(\omega_2^2+(\omega_3^2-\omega_2^2)r_2^2+\frac{\alpha\nu^2}{1-\alpha}\right),
           \end{split}
           \label{mainequation}
        \end{equation}
        which can also be written as
        \begin{equation}
        \begin{split}
            (1-\alpha_2^2)^2r_2^{\prime 2}=\frac{1}{r_2^2}&\Bigg[\left((1+\alpha_2^2)\omega_0^2\alpha-\omega_2^2-\frac{\alpha(\nu^2-v^2)}{1-\alpha}\right)r_2^2(1-r_2^2)\\&-(\omega_3^2-\omega_2^2)r_2^4(1-r_2^2)+v_2^2+(v_1^2-v_2^2)r_2^2\Bigg].
            \end{split}
        \end{equation}
        It can be seen above that the differential equation in $r_2$ has three zeroes corresponding to the turning points $(r_2^{'}=0)$. We want to look for spiky string solution which demands solution with two turning points. At large $r_2$ limit, the above equation describes a string which does not reach the boundary under the condition $\omega_3^2<\omega_2^2$. Looking at the expression for roots, we can take one of the roots to be $r_2=1$ resulting in fixing the value of one of the constants of motion i.e, $v_1=0$. Substituting the value of $v_1$, the above equation reduces to
        \begin{equation}
           (1-\alpha_2^2)^2r_2^{\prime 2}  =\frac{1-r_2^2}{r_2^2}\Bigg[\left((1+\alpha_2^2)\omega_0^2\alpha-\omega_2^2-\frac{\alpha(\nu^2-v^2)}{1-\alpha}\right)r_2^2-(\omega_3^2-\omega_2^2)r_2^4+ v_2^2\Bigg].
        \end{equation}
        Here we can see that again we get two zeroes, where $r_2=1$ is a double zero. Substituting this value of $r_2$ in the above equation, the right hand side of the equation reduces to $(1+\alpha_2^2)\omega_0^2\alpha=\omega_3^2+\cfrac{\alpha(\nu^2-v^2)}{1-\alpha}- v_2^2$ and using $v_1=0$ in equation (\ref{secondconstraint}), we find the value of $\alpha_2$ to be $\left(\cfrac{-\omega_3v_2-\nu v\frac{\alpha}{1-\alpha}}{\alpha\omega_0^2}\right)$. Putting the value of $\alpha_2$ in the above equation we get a quadratic equation in $\alpha\omega_0^2$, the roots of which is complicated and does not give simplified solution to our equation. Hence we will drop this line of action and take a different approach in order to solve the problem. We want to express  (\ref{mainequation}) in terms of variable $y$, where $y$ can be defined as
        \begin{equation}
         y=\frac{1}{1-2r_2^2}=\frac{1}{\cos2\beta_1}.
        \end{equation}
        Here we have used $r_2=\sin\beta_1$ where $\beta_1$ is the global coordinate of $S^3$. Substituting the value of $r_2$, we get the equation of the form
        \begin{equation}
        \begin{split}
          (1-\alpha_2^2)^2&y^{' 2}=2y\Bigg[4y^2v_1^2(y-1)+4y^2v_2^2(y+1)-\omega_3^2(y-1)^2(y+1)\\&-\omega_2^2(y+1)^2(y-1)+2y(y^2-1)\left(\alpha\omega_0^2(1+\alpha_2^2)-\frac{\alpha(\nu^2-v^2)}{1-\alpha}\right)\Bigg],  
          \end{split}
        \end{equation}
      which can be written in terms of Polynomial as
        \begin{equation}
         y^{\prime}=\frac{\sqrt{2yP(y)}}{1-\alpha_2^2},
         \end{equation}
        where the expression for the polynomial $P(y)$ is given as
        \begin{equation}
        \begin{split}
       &P(y)=y^3\left[4v_1^2+4v_2^2-\omega_3^2-\omega_2^2+2\left(\alpha\omega_0^2(1+\alpha_2^2)-\frac{\alpha(\nu^2-v^2)}{1-\alpha}\right)\right]+\omega_2^2-\omega_3^2\\&+y^2\left[-4v_1^2+4v_2^2+\omega_3^2-\omega_2^2\right]+y\left[\omega_2^2+\omega_3^2-2\left(\alpha\omega_0^2(1+\alpha_2^2)-\frac{\alpha(\nu^2-v^2)}{1-\alpha}\right)\right]
       \end{split}
       \end{equation}
      \begin{equation}
       \equiv \left[4v_1^2+4v_2^2-\omega_3^2-\omega_2^2+2\alpha\left(\omega_0^2(1+\alpha_2^2)-\frac{(\nu^2-v^2)}{1-\alpha}\right)\right](y-y_1)(y-y_2)(y-y_3),
      \end{equation} 
      where $y_1,y_2$ and $y_3$ are considered as the three roots of the polynomial $P(y)=0$. Consistent string solution demands all the roots of the polynomial to be real. Here we can reduce the number of constants by eliminating some of them in terms of the other using the constraint equations.\\
      Let us take the polynomial $P(y)$ to have one negative and two positive real roots. From the algebra of the third order polynomial, we can write the product of the roots determined by the parameter as
       \begin{equation}
           a\equiv 4v_1^2+4v_2^2-\omega_3^2-\omega_2^2+2\alpha\left(\omega_0^2(1+\alpha_2^2)-\frac{(\nu^2-v^2)}{1-\alpha}\right)=\frac{\omega_3^2-\omega_2^2}{y_1y_2y_3}
       \end{equation}
       Here we can see that $`a'$ is negative provided some of the constants in the expression i.e, $v_1,v_2,v,\alpha,\alpha_2$ are small. This condition results in the polynomial $P(y)$ being positive between the two positive roots taken. Now, we have to fix the two physical constants $y_1$ and $y_2$, in between which our polynomial is positive.
       \begin{equation}
           P(y)=\frac{\omega_3^2-\omega_2^2}{y_1y_2y_3}(y-y_1)(y-y_2)(y-y_3)=8v_2^2\frac{(y-y_1)(y-y_2)(y-y_3)}{(1-y_1)(1-y_2)(1-y_3)}.
       \end{equation}
       By using above expression, we can write $y_1$ in terms of $y_2$ and $y_3$ as,
       \begin{equation}
           y_1=-\frac{y_2y_3}{y_2+y_3+y_2y_3\frac{\omega_2^2+\omega_3^2-2\left(\alpha\omega_0^2(1+\alpha_2^2)-\frac{\alpha(\nu^2-v^2)}{1-\alpha}\right)}{\omega_2^2-\omega_3^2}}.
       \end{equation}
       The expressions for $v_1$ and $v_2$ in terms of the roots $y_1,y_2,y_3$ are,
       \begin{equation}
           v_2^2=\frac{\omega_3^2-\omega_2^2}{8}\frac{(1-y_1)(1-y_2)(1-y_3)}{y_1y_2y_3},~~v_1^2=\frac{\omega_3^2-\omega_2^2}{8}\frac{(1+y_1)(1+y_2)(1+y_3)}{y_1y_2y_3}
       \end{equation}
       Here we clearly see that for the range of the solution $-1\leq y_1\leq 0\leq y_2\leq y_3\leq 1$, we get $v_1^2>0$ and $v_2^2>0$ which confirms the consistency of our roots for the polynomial $P(y)$. Here in this article, since we want to study the $N$-spike string, we need to glue together $2n$ number of integrals between $y_2$ and $y_3$. So we replace $\int d\xi$ by
       \begin{equation}
          \int d\xi= 2n\int_{y_2}^{y_3}\frac{dy}{y'}=\frac{2n (1-\alpha_2^2)}{\sqrt{-2a}}F_1,
       \end{equation}
       where $F_1$ is the integral which has been calculated in the appendix. Now we will find the winding number $m$ defined in (\ref{winding}) as,
       \begin{equation}
           m=\frac{v+\alpha_2\nu}{\pi \sqrt{-2a}}nF_1.
           \label{m}
       \end{equation}
       We want to calculate the equation for $\nu$ which can be obtained by solving for $v$ and using the second virasoro constraint (\ref{secondconstraint}) as
       \begin{equation}
           \omega_2v_1+\omega_3v_2+\frac{\nu \alpha}{1-\alpha}\left(\frac{\pi m \sqrt{-2 a}}{nF_1}-\alpha_2\nu\right)=-\alpha_2\omega_0^2\alpha
       \end{equation}
       We can eliminate one of the constant by using the condition given in (\ref{zerowinding}) as
       \begin{equation}
           -2v_2F_6+\alpha_2\omega_3 F_1=0.
           \label{alpha2}
       \end{equation}
       The above relation can be used to calculate $\alpha_2$ which in turn can be substituted to find the winding number done later in this section.\\ The expression for conserved charges in terms of Jacobi Elliptic integrals are as follows
       \begin{equation}
       \begin{split}
       \frac{\alpha\pi\mathcal{J}_1}{n}&=\frac{v_1\alpha_2}{\sqrt{-2a}}F_1+\frac{\omega_2}{2\sqrt{-2a}}F_3,~~~~~\frac{\alpha\pi\mathcal{J}_2}{n}=\frac{v_2\alpha_2}{\sqrt{-2a}}-\frac{\omega_3}{2\sqrt{-2a}}F_2\\&\frac{\pi\mathcal E}{n}=\frac{\omega_0(1-\alpha_2^2)}{\sqrt{-2a}}F_1,~~~~~\frac{(1-\alpha)\pi\mathcal J}{n}=\frac{(v\alpha_2+\nu)}{\sqrt{-2a}}F_1.
       \end{split}
       \end{equation}
       where $F_1,F_2,F_3$ are the elliptic integrals defined in the appendix and $$E=2\pi T\mathcal{E},~~ J=2\pi T\mathcal{J}, ~~ J_1=2\pi T\mathcal{J}_1,~~ J_2=2\pi T\mathcal{J}_2 $$ has been used in the above equations. Now we would like to speculate the configuration of the rigid $N$-spike string when we generalise the motion of the string to the addition of angular momentum $J$ and winding in $S^3$. 
        This can be checked by evaluating the derivative at maximum value of $\beta_1$ or minimum value of $y=y_2$ at fixed time. At a given fixed time we can also  calculate the number of spikes using the relation $\Delta {\phi_2}=\frac{2\pi}{2n}$, where $\Delta \phi_2$ is the angle between a valley or a minimum and a maximum. We get $\phi_2$ from the initial coordinate to be, 
        \begin{equation}
           X_1=\frac{R}{\sqrt{\alpha}}\cos{\beta_1}e^{i\phi_2},~~~ \phi_2=\omega_2\tau+\int d\xi f_1^{'}.
       \end{equation}
       which in turn gives the expression for $\Delta \phi_2$ as 
       \begin{equation}
           \Delta\phi_2=\int d\phi_2=\frac{1}{\alpha_1^2-\alpha_2^2}\int d\xi \left(\frac{v_1}{1-r_2^2}-\frac{\omega_2}{\omega_3}\frac{v_2}{r_2^2}\right)=\frac{2}{\sqrt{-2a}}\left(v_1F_5+\frac{\omega_2}{\omega_3}v_2F_6\right).
       \end{equation}
       Now we will evaluate the derivative at the minimum value of $y=y_2$. This gives us the derivative at the maximum value of $\beta_1$ at a fixed time
       \begin{equation}
           \frac{d\beta_1}{d\phi_2}\bigg|_{y=y_2}=\frac{\beta_1^{'}d\xi}{\omega_2d\tau+f_1^{'}d\xi}\bigg|_{y=y_2}
       \end{equation}
       Using that for fixed time i.e., $dt=\omega_0 d\tau=0$, gives the the following
       \begin{equation}
           \frac{d\beta_1}{d\phi_2}\bigg|_{y=y_2}=\frac{\beta_1^{'}}{f_1^{'}}\bigg|_{y=y_2}
       \end{equation}
       Solving the above equation we get,
       \begin{equation}
            \frac{d\beta_1}{d\phi_2}\bigg|_{y=y_2}=\frac{P(y)}{{\sqrt{2}y^{\frac{3}{2}}\sqrt{y-1}}}\frac{
            \sqrt{y+1}}{2v_1+\alpha_2\omega_2}\frac{1}{\left(\cfrac{\alpha_2\omega_2}{2v_1+\alpha_2\omega_2}+y\right)}\Bigg|_{y=y_2}.
       \end{equation}
       We can clearly see here that for $y=y_2$, the polynomial $P(y_2)$ which is there in the numerator vanishes but in general the denominator does not. The denominator vanishes only under the condition $y_2=-\frac{\alpha_2\omega_2}{2v_1+\alpha_2\omega_2}$ but this corresponds to motion only in $S^3$, also this gives negative $y_2$ which is not a consistent choice. Thus we observe that in general when we extend the motion of the string to $S^3$, the spikes get rounded off at $y=y_2$ and not end in cusps. Similar argument has been presented in \cite{Ishizeki:2008tx}. \\
        We will now calculate the expression for the winding number. This can be done by solving for $\alpha_2$ in (\ref{alpha2})  and replacing in (\ref{m}), we get
       \begin{equation}
           m=\frac{\nu}{2\pi\omega_3}nF_6\sqrt{(1-y_1)(1-y_2)(y_3-1)}
       \end{equation}
       It can be seen from the expression that the winding number depends on four independent parameters $n, y_1,y_2,y_3$. All other parameters can be expressed in terms of these the independent parameters. Given the number of independent parameters, the energy can be written as 
       \begin{equation*}
           \mathcal{E}=\mathcal{E}(\mathcal{J}_1,\mathcal{J}_2,\mathcal{J},n,m)
       \end{equation*}
       Due to the complicated form of the roots further assumptions needs to be taken to solve this. We can check the similar nature of the rounding off of the spikes when we introduce some flux to the background.
            \section{Conclusion and outlook}
            We have elucidated the emergence of classically integrable NR model underlying in the two dimensional string-sigma model that describes the mixed flux-supported $AdS_3\times S^3_1\times S^3_2\times S^1$ probed with fundamental string. We have emphasized on the generic rotating string ansatz during the NR formalism in our present scenario. The parameter $\alpha$ of the relative geometry of the spheres in the background has played considerably significant role in the overall construction of the integrable model. In support of our construction, we reproduced the Lagrangian and the independent Uhlenbeck constants for our system and showed that all these pieces justifiably assume quite similar forms as those of the undeformed general NR system. What is exclusive here is that the Lagrangian includes some extra terms which involve the flux parameters as well as $\alpha$. 
            Nevertheless, the integrals of motion are not affected anyway by the parameter $\alpha$ and they individually satisfy the constraints of the geometries in $AdS_3\times S^3_1\times S^3_2$. Moreover, the condition for Liouville integrability is also satisfied by the integrals of motion of the system we developed. After looking at the integrable structure of the system, we study the rigidly rotating spiky string solutions with energy $E$ and angular momentum $J$. The string here is rotating in pure $S^3\times S^1$ with the addition of non-zero angular momentum $J$ in $S^1\subset S^3$. We find the expression for the conserved charges as well as winding number $m$, in terms of elliptic integrals. Finally, we mathematically speculate the configuration of the spiky string profile. We establish that, due to the addition of non-zero angular momentum $J$ in $S^1$, the spikes of the rigidly rotating $N$ spike string, get rounded off and not end in cusp. The generalization of the NR solutions to a class of pulsating string should be interesting as the pulsating strings present a more stable family of solutions than the usual rotating ones. It is well-known that the stability of string solutions for definite spin values increases with the increase in the number of pulsations\cite{Khan:2005fc}. We expect that string pulsating in both $S^3_1$ and $S^3_2$ simultaneously with two linearly dependent oscillation numbers in each sphere can acquire improved stability than those in only one sphere for a specific chosen value of spin along AdS directions. We wish to come back to this problem in near future. 
          \section*{Appendix A}
          Some of the useful integrals used are summarized below
          \begin{equation}
          \begin{split}
              F_1=\int_{v_2}^{v_3}&\frac{dy}{\sqrt{-y(y-y_1)(y-y_2)(y-y_3)}},F_2=\int_{v_2}^{v_3}\frac{dy(1-y)}{y\sqrt{-y(y-y_1)(y-y_2)(y-y_3)}}\\&F_3=\int_{v_2}^{v_3}\frac{dy(1+y)}{y\sqrt{-y(y-y_1)(y-y_2)(y-y_3)}}
              \end{split}
          \end{equation}
          The above integrals takes the form of the elliptic integrals as follows
          \begin{equation}
          \begin{split}
              &F_1=\frac{2}{\sqrt{y_3(y_2-y_1)}} K\left[\sqrt{\frac{y_1(y_2-y_3)}{y_3(y_2-y_1)} }\right],\\&F_2=\frac{-2}{y_1y_2}\sqrt{\frac{y_2-y_1}{y_3}}E\left[\sqrt{\frac{y_1(y_2-y_3)}{y_3(y_2-y_1)} }\right]+\frac{2(1-y_1)}{y_1\sqrt{y_3(y_2-y_1)}}K\left[\sqrt{\frac{y_1(y_2-y_3)}{y_3(y_2-y_1)} }\right],\\&F_3=\frac{-2}{y_1y_2}\sqrt{\frac{y_2-y_1}{y_3}}E\left[\sqrt{\frac{y_1(y_2-y_3)}{y_3(y_2-y_1)} }\right]+\frac{2(1+y_1)}{y_1\sqrt{y_3(y_2-y_1)}}K\left[\sqrt{\frac{y_1(y_2-y_3)}{y_3(y_2-y_1)} }\right]
              \end{split}
          \end{equation}
          Some other integrals used in the manuscript are:
          \begin{equation}
              F_4(y_2,y)=\int_{y_2}^y dy \frac{y}{\sqrt{-y(y-y_1)(y-y_2)(y-y_3)}}
          \end{equation}
          \begin{equation}
             F_5(y_2,y)=\int_{y_2}^y dy \frac{y}{(1+v)\sqrt{-y(y-y_1)(y-y_2)(y-y_3)}} 
          \end{equation}
          \begin{equation}
              F_6(y_2,y)=\int_{y_2}^y dy \frac{y}{(1-y)\sqrt{-y(y-y_1)(y-y_2)(y-y_3)}}
          \end{equation}
         On solving the above integrals we get elliptic functions 
         \begin{equation}
             F_4(y_2,y_3)=\frac{2y_2}{\sqrt{y_3(y_2-y_1)}}\Pi\left[\frac{y_3-y_2}{y_3},\sqrt{\frac{y_1(y_2-y_3)}{y_3(y_2-y_1)}}\right]
         \end{equation}
          \begin{equation}
             F_5(y_2,y_3)=\frac{2y_2}{(1+y_2)\sqrt{y_3(y_2-y_1)}}\Pi\left[\frac{y_3-y_2}{y_3(1+y_2)},\sqrt{\frac{y_1(y_2-y_3)}{y_3(y_2-y_1)}}\right]
         \end{equation}
             \begin{equation}
             F_6(y_2,y_3)=\frac{2y_2}{(1-y_2)\sqrt{y_3(y_2-y_1)}}\Pi\left[\frac{y_3-y_2}{y_3(1-y_2)},\sqrt{\frac{y_1(y_2-y_3)}{y_3(y_2-y_1)}}\right]
         \end{equation}

\end{document}